\documentclass[aps,prl,showpacs,twocolumn,lengthcheck]{revtex4}

\usepackage{graphicx}
\usepackage{amssymb,latexsym,amsthm}

\begin{document}

\title{Sparse repulsive coupling enhances synchronization in complex networks}
\author{I. Leyva}
\author{I. Sendi\~na--Nadal}
\author{J.A. Almendral}
\author{M.A.F. Sanju\'an}
\affiliation{Dpto. de Ciencias de la Naturaleza y F\'{\i}sica
  Aplicada, Universidad Rey Juan Carlos, c/ Tulip\'an s/n,   
28933 M\'ostoles, Madrid, Spain}

\date{\today}

\begin{abstract}
Through the last years, different strategies to enhance synchronization in complex networks 
have been proposed. In this Letter, we show that the synchronization in a small-world network
of attractively coupled non-identical neurons is strongly improved by
adding a tiny fraction of phase-repulsive couplings. 
By a purely
topological analysis that does not depend on the dynamical model, we link the emerging dynamical
behavior to the structural properties of the sparsely coupled repulsive network. 
\end{abstract}
\pacs{87.19.La, 05.45.Xt }
\keywords{}

\maketitle

Synchronous oscillations in a large ensemble of oscillators are considered as one of the mechanisms
in biological networks to transmit and code information, especially in the brain \cite{phisio1}. 
Recent experiments have pointed out the important role that the complex structure of 
connectivity has in this collective behavior \cite{BenJacobPRL02}, obtaining the
signature of an underlying small-world (SW) network  by indirect measures in neuronal culture 
samples \cite{JiaPRL04} or using functional magnetic resonance imaging in 
humans \cite{EguiluzPRL05}. Theoretically,
several strategies have been developed with the aim of finding the best 
way to achieve synchronization in complex networks
\cite{BarahonaPRL02,StrogatzNAT01,BarabasiRMP02}.
These approaches have mainly focused on the role that weighted links play in
heterogeneous networks \cite{NishikawaPRL03,MotterPRE05,BoccalettiPRL05}, shortest paths between nodes 
and clustering structure in SW networks \cite{LagoPRL00},
or the input degree each node receives regardless of the net structure \cite{HaslerPRL05}. 

Most of this research has been devoted to attractively coupled dynamical elements. 
However, it is known that biological networks
combine different types of connections to improve synchronization
and  transmission performance, as in the case of the coexistence of
excitatory and inhibitory synapses in the brain \cite{AbarbanelRMP}.
Nevertheless, little attention has been paid to the effect of repulsive
 coupling, or to the interaction between different types of
 coupling. The scarce literature addressing
synchronization in repulsively coupled oscillators considers global
or local coupling \cite{MossPRE01,TsimringPRL05}, but the influence
of the network structure is still an open question.

In addition, almost all the published work on synchronization in complex networks basically
deals with arrays of identical units.
However, heterogeneity of elements is an inherent feature present in natural 
systems which can be especially relevant in the dynamics of biological networks \cite{Aradi02}. 
In this Letter, we explore the influence of the network topology on the dynamics of  
non-identical coupled units, when a small fraction of the links is phase-repulsive.  
We show that sparse repulsive links in a SW structure can induce 
a coherent oscillatory state when the equivalent SW composed of only attractive connections is not able to
synchronize or even to activate the ensemble. Then, just by means of an
analysis focused on the connectivity matrix, we link the emerging dynamical
behavior to the structural properties of the sparsely coupled repulsive network. 

We study the dynamics of 
an ensemble of non-identical locally coupled Hodgkin-Huxley (HH) 
neurons \cite{HodgkinJPHYSIOL52} considered as
spatially isopotential cells
\begin{eqnarray}  
C \dot{V_i}&=& I_i - I_{i}^{ion}(V_i,x_i) +d\sum_{j} \hat{L}_{ij}V_j \label{HH} \\
\dot{x_i}&=&\alpha_x(1-x_i)-\beta_x x_i.\nonumber
\end{eqnarray}                                                 

\noindent
Here, $V_i$ is the voltage across the membrane of neuron $i$ of capacitance $C$, and 
$\dot x =\{\dot m,\dot n,\dot h\}$ describes the gating of the ion channels. 
$I_{i}^{ion}=g_{Na}m_i^3h_i(V_i-V_{Na})+ g_K n_{i}^{4}(V_i-V_K)+g_l(V_i-V_l)$ is the 
ionic current mainly carried by Na$^+$ and K$^+$ ions and other ionic currents
through voltage dependent channels. 
These currents are driven by the voltage difference 
with respect to the equilibrium potentials $V_{Na}$, $V_{K}$ and $V_{l}$ and the maximal 
ionic conductances $g_{Na}$, $g_{K}$ and $g_l$. 
Functions $\alpha_{x}$ and $\beta_{x}$  are voltage-dependent 
rate constants.
Parameter values and functions are the standards in the literature \cite{HodgkinJPHYSIOL52,parametros}. 
$I_i$ accounts for any external bias current, which has been chosen as a 
control parameter to introduce heterogeneity in the population by setting
$I_i$ uniformly distributed within the interval $I_0\pm\Delta I$. The value 
$I_0=9\mbox{ $\mu$A/cm$^2$}$ is fixed close to the point where an inverse Hopf bifurcation
occurs, $I_b=9.5\mbox{ $\mu$A/cm$^2$}$. This way, for the chosen $\Delta I=0.2$,
about $90\%$ of the neurons stay around the silent state,
$V_{\mbox{\scriptsize{rest}}}=-65\mbox{ mV}$, 
while the rest will fire periodically.

\begin{figure}
  \centering  \includegraphics[width=0.49\textwidth]{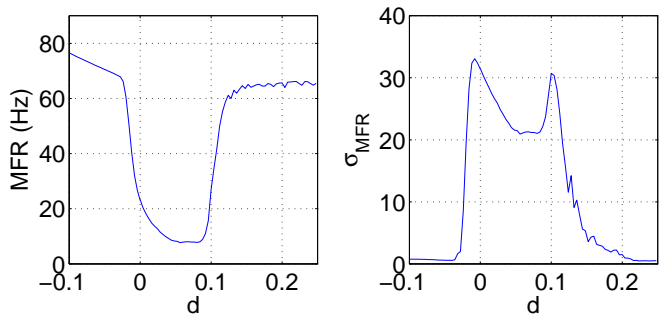}
  \caption{Mean firing rate (left) and its standard deviation (right) for a $N=400$  
regular array as a function of $d$, each point averaged over 100
realizations. The negative sign of $d$ comes from
the fully phase-repulsive connection matrix.}\label{local-coupling}
\end{figure}

The coupling structure in ~(\ref{HH}) is given by
$\hat{L}_{ij}=\,L_{ij}/k_i$, where $L_{ij}$ is the Laplacian matrix \cite{BoccalettiRMP},
$k_i$ normalizes the connection strength by the number of incoming
links to node $i$, and the coefficient $d$ stands for the global
coupling strength. 

\indent {\it Local coupling.\textemdash}
Initially we consider the effect of a regular lattice 
on an ensemble of $N$ neurons, both for fully phase-attractive and
phase-repulsive coupling. The Laplacian matrix for the first case is
$L_{i,i\pm 1}=1$, $L_{ii}=-2$ and $L_{ij}=0$ otherwise. And, for the
second one, it is $L_{i,i\pm 1}=-1$, $L_{ii}=2$ and $L_{ij}=0$ otherwise.

Figure~\ref{local-coupling} shows the global mean firing rate (MFR) and 
its standard deviation $\sigma_{\mbox{\scriptsize{MFR}}}$ as a function of $d$ ranging
from negative to positive values.  The negative sign of $d$ comes from
the fully phase-repulsive connection matrix. 
When $d>0$ is large enough, the system is frequency entrained to a phase
synchronization state.  Equivalently, for a sufficient $d<0$, the
system reaches an anti-phase synchronization state. It can be noted from
Fig.~\ref{local-coupling} that the entrainment with negative
couplings is achieved for smaller absolute values of $d$ compared to
the case with positive ones.
This indicates that a phase-repulsive coupling is more effective to activate
and entrain the whole network. Many biological systems exhibit this
kind of repulsive coupling when
their dynamical units are in competition with each other. Known
examples are the inhibitory coupling present in neuronal circuits 
associated to a synchronized behavior in central pattern generators \cite{KopellSIAM94}  or calcium
oscillations in epileptic human astrocyte  cultures \cite{MossCHAOS03}.  

\indent {\it Non local random coupling.\textemdash}
Our main interest is to explore the influence of a SW-like connection
topology in the activation and synchronization of the network.  
From the results obtained in the previous section, we know that 
a small positive coupling strength is 
less efficient than a negative coupling to activate and synchronize
the whole network.
Taking this into account, we consider
now the possibility of non-local couplings both positive and
repulsive. 
The global coupling strength is fixed to $d=0.1$, i.e., within 
the unsynchronized regime for local positive coupling as shown in
Fig.~\ref{local-coupling}. The Laplacian matrix $L$ is modeled now by keeping the 
regular short-range connections positive, $L_{i,i\pm 1}=+1$, and by randomly adding
(rather than rewiring) a fraction $p$ of the $(N-1)(N-2)/2$ possible
long-range links, being negative with probability $p_{n}$.

\begin{figure*}
\centering \includegraphics[width=0.8\textwidth]{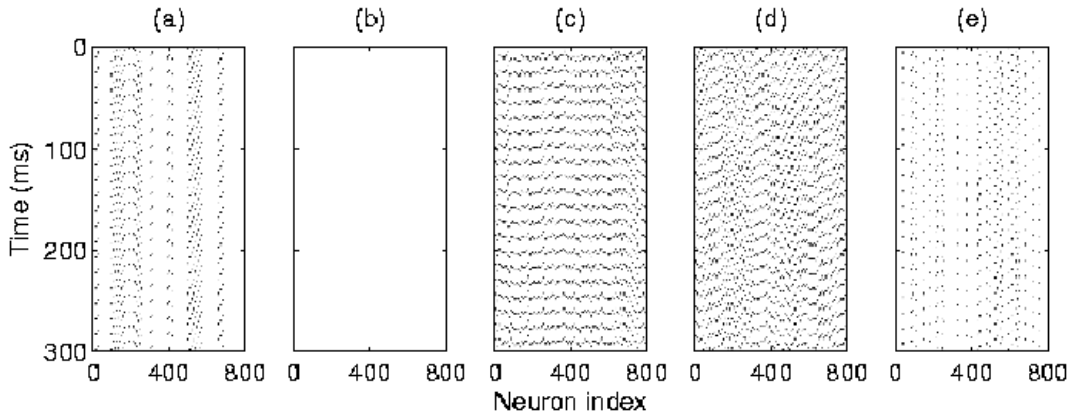}
\caption{Space-time plots of the neuron voltage for a $N=800$ HH units
  network, with $\Delta I=0.2$, $d=0.1$, and different coupling
  connectivities: 
(a) Local coupling with $p_{n}=0$;
(b) network with long range couplings, $p=p_c=0.0055$, and $p_{n}=0$;
(c) $p=p_c=0.0055$ and $p_{n}=0.3$;
(d) $p=p_c=0.0055$ and $p_n=0.45$;
(e)  $p=0.015$ and $p_n=0.3$.}
\label{rasters} 
 \end{figure*}

Figure~\ref{rasters} shows space-time plots of the voltage variable through the whole array for different
probabilities $p$ and $p_{n}$. As expected, in the absence of
long-range connections, few more than the initial $10\%$ of the neurons
is firing for the chosen coupling strength $d$, i.e., the array
is not even activated as shown in Fig.~\ref{rasters}(a). When long-range
links are included, the first observation
is that for any $p$, a minimum fraction of the new added links needs
to be repulsive in order to increase 
the activity of the network. This becomes evident when comparing Fig.~\ref{rasters}(b) with Figs.~\ref{rasters}(c)-(e). 
In  Fig.~\ref{rasters}(b) the activity generated by the $10\%$ of
initially active neurons is 
reduced, or even annihilated, when all long-range connections are attractive ($p_n=0$). However,
the scenario completely changes when, for the same $p$, some of the shortcuts are
repulsive ($p_n>0$) as in Figs.~\ref{rasters}(c)-(e) where self-sustained 
electrical activity emerges for nonzero $p_{n}$.  
In addition, we observe the existence of optimal probabilities $p$
and $p_n$ for which the collective oscillation becomes maximally
phase-coherent. This fact can be observed by comparing  Fig.~\ref{rasters}(c),
where $p$ and $p_n$ are optimal, and Fig.~\ref{rasters}(d) where $p_n$ is the same but
$p$ is slightly higher.  

To study quantitatively how the dynamics is affected by $p$ and $p_{n}$, 
we measure the MFR of the network and the standard deviation 
of the global electrical voltage, $V(t)=\sum_{i=1}^N V_i(t)$, obtained
as $\sigma_{V}=\sqrt{\langle V^2(t)\rangle-\langle V(t)\rangle^2}$. 
While the MFR gives us an estimation of how much the network is
activated, the $\sigma_{V}$ defines how 
coherent is the activity of the entire network. If the network is
fully activated, the MFR approaches to a rate of around
70 Hz, whereas $\sigma_{V}$ is maximal if this activity is synchronized. 

\begin{figure}
\centering\includegraphics[width=0.49\textwidth]{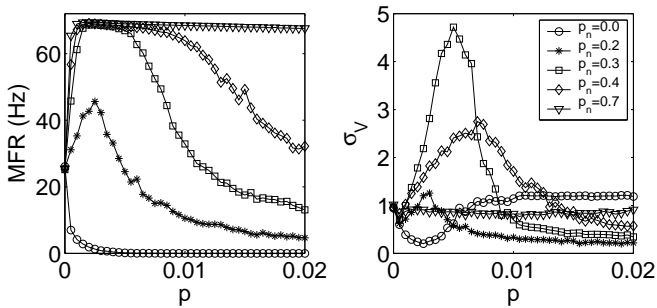}
\caption{MFR (left) and network coherence $\sigma_{V}$ (right) as a function of $p$ 
for several $p_{n}$ in a $N=800$ network. Each point is averaged over 100 simulations, 1 s long
(transients avoided), for different network and initial conditions
realizations. Note that the legend applies to both figures.}
\label{quantitatively}
\end{figure}

We have plotted in Fig.~\ref{quantitatively} both the MFR and the
$\sigma_{V}$ as a 
function of the probability $p$ for
different values of $p_{n}$. The signature of a network resonance both in $p$ and
$p_{n}$ is clear from this figure: 
the frequency entrainment increases and phase synchronization is maximally
enhanced for the optimal values $p=p_c$ and $p_n\approx 0.3$. The probability $p_c$
depends slightly on $p_n$, shifting to higher $p$ as $p_n$ increases,
but remaining very small. 
The interplay between topology and dynamics 
becomes evident when we observe that the critical link probability depends strongly on the 
size ensemble as $p_c\propto \ln(N)/N$, i.e., coincides with the birth of the giant 
connected component (GCC) when only the random part of the network is
considered. 

Recently \cite{BoccalettiPRL05,BorjaChaos}, the method of the
\emph{master stability  function} \cite{PecoraPRL80} 
has been successfully
used to analyze whether the network structure has some bearing on the
dynamics evolving on it. 
However, this approach requires the dynamical units to be identical
(which is not our case) and, generally, the results are model-dependent. 
Therefore, in order to understand 
the influence of a complex connectivity, we use a purely structural analysis based 
on the properties of the matrix $\hat{L}$. To this purpose,
we ignore the intrinsic dynamics of the neurons in Eq.~(\ref{HH}), 
that is, we just consider $\dot{{\bf V}}=d\hat{L}{\bf V}$. 
Then, there is a basis in which $V_i \approx \exp(d\lambda_i t)$, where
$\lambda_i$ are the eigenvalues of $\hat{L}$. 
It is well known that all the eigenvalues of the Laplacian associated to a 
network with only attractive couplings are negative. 
However, when we add some repulsive connections, $\hat{L}$ has
positive and negative eigenvalues.
We find that any set of initial states rapidly 
evolves into the subspace $S^+$ associated to the positive eigenvalues
within a time smaller than the characteristic temporal scale of the
neuronal dynamics ($\tau\approx 15$ ms).

To quantify the effect of $S^+$, we note that, for a given positive $\lambda_i$, $e^{d\lambda_i}$ is a
measure of how much the system spreads into the subspace defined by
the corresponding eigenvector. Then, the ratio $e^{d\lambda_i
  t}/e^{d\lambda_{\mbox{\scriptsize{max}}}t}=e^{d(\lambda_i -
  \lambda_{\mbox{\scriptsize {max}}})t}$ measures how
different is the evolution in that subspace with respect to the
one where the system develops faster. By defining the geometric average 
$g(t)= \prod_{i=1}^N e^{d(\lambda_i -\lambda_{\mbox{\scriptsize{max}}})t/N} =
e^{d(\langle{\lambda}\rangle - \lambda_{\mbox{\scriptsize{max}}})t}$, we can
estimate the homogeneity of the evolution in $S^+$ with a number in $(0,1]$. A
value close to $1$ means the system evolves similarly in all dimensions of $S^+$, whereas a low $g$
implies that its behavior is determined by those vectors with the largest associated eigenvalues.

\begin{figure}
\centering \includegraphics[width=0.49\textwidth]{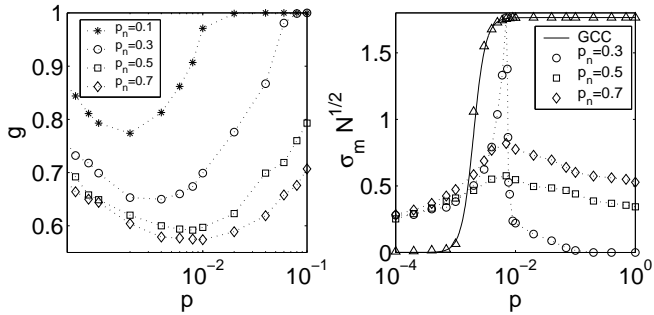}
\caption{
Left: Dependence of $g$ with the adding
  probability $p$, in a log--linear scale, 
for different probabilities $p_n$. Each point
is an average over $100$ different realizations of a $N=800$ network. 
Right: Deviation $\sigma_m\sqrt{N}$ of the mean state {\it vs.} $p$ for different $p_n$ values in a 
log--linear scale, with $N=800$.
Each point averages $1000$ runs after a transient of $100$
iterations and fixed $p_{\_}=0.1$. 
\label{defor-pico}}  
\end{figure}

We are interested in the behavior of $g(t)$ as a function of $p$ and
$p_n$. As the shape of $g(t)$ with $p$ is not very sensitive to time,
we fix $t=d^{-1}\sim\tau$ to focus our study within the time scale of
our dynamical unit (see Fig.~\ref{defor-pico}(left)). We observe that
$g=g(\tau)$ 
presents a minimum at $p_c$ which is lower
for higher values of $p_n$, 
and whose position shifts to higher $p$ as $p_n$ increases, as
observed in the numerical simulations (Fig.~\ref{quantitatively}(right)). 
This means that, for values of $p$ far from $p_c$, i.e. where
$g\approx 1$, the global dynamics is basically determined  by only one 
positive eigenvalue, ${\bf V}(t) = {\bf V}_0 \exp(\lambda t)$. On the
contrary, for values of $p$ close to $p_c$, we need to consider not just one but several
eigenvalues (the largest ones) to account for the global
dynamics. 
Therefore, the intrinsic dynamics of the system is minimally
constrained by the structure that arises around $p_c$ due to the repulsive shortcuts.

To analyze if the previous structural result affects other dynamics imposed on it, we consider a
discrete spin-like dynamics in which each node $i$ has only two possible states $s_i= \pm 1$, with
a probability $p_{\_}$  of being $s_i=-1$. Consequently, with the
same link matrix $\hat{L}$ 
studied above, node $i$ receives an input $h_i = \sum_j \hat{L}_{ij} s_j \in [-2,2]$. 
Hence, as other authors have pointed out ~\cite{watts99,zhou05}, 
these spin-like networks can be regarded as a pattern of the internal states and their evolution represent the 
global dynamics.
Notice that the neighbor
vertices linked repulsively contribute to the input with the opposite
state. Then, in this model it is implicit that nodes linked with an attractive
connection tend to follow the same evolution, 
whereas repulsive connection leads them to evolve differently.

We can prove analytically that the distribution 
of $h_i$ presents two peaks: $\mu_1 = -2p_{\_}\sqrt{1-4p_n(1-p_n)}$
and $\mu_2=2(1-p_{\_})\sqrt{1-4p_n(1-p_n)}$. 
Then, we choose to evolve the network according to the following local
majority rule: the new state of node $i$ is updated to $s_i(n+1) = +1$ if $h_i(n) > \mu_2$, $s_i(n+1) = -1$ if
$h_i(n) < \mu_1$ and $s_i(n+1) = s_i(n)$ otherwise (i.e., the vertex
keeps its state).

When we average the mean state of nodes after a transient, we find
that the system changes its behavior at $p \approx p_c$. However, since these changes in the average
depend on $p_n$ and $p_{\_}$, we focus our attention in the deviation $\sigma_m$ of the mean state. This quantity
measures how many different states are allowed for that particular network structure. It can be seen in
Fig.~\ref{defor-pico}(right) that the maximum of $\sigma_m$ is reached 
when the GCC associated to the long-range links spans the whole network with a minimal number of links.

This shows how $p$ and $p_n$ contribute to improve the
synchronization even for this discrete dynamics. When the net is essentially a lattice, 
the system remains in disorder since there are few $\lambda_i >0$ to
spread the activity throughout. If we have a fully connected network with many $\lambda_i >0$, the
whole system is activated but, since all dimensions in $S^+$ contribute similarly to the dynamics, the topology
constrains the neurons to evolve alike when they have different intrinsic dynamics. On the contrary, the
structure close to $p_c$, due to the presence of phase-repulsive links 
is such that, not only the activity is enhanced, but also the topology is compatible
with the heterogeneity of the system.

In summary, we have shown how a small fraction of phase-repulsive links
can enhance  synchronization in a complex network of dynamical units. A structural analysis allows us to 
obtain information about how the topology influences the
dynamics. Surprisingly, around $p_c$, the versatility arising from the
network structure due to $p_n$ drives the system to a more ordered
state, while far from $p_c$ the stiffness of the structure freezes the initial disorder. 

We thank Dr.~J.R. Pel\'aez for fruitful discussions. 
This work has been financially supported by the Spanish Ministry of
Science and Technology under project No. BFM2003-03081 and
the URJC project No. PPR2004-04. All numerical computations
were performed at the Centro de Apoyo Tecnol\'ogico of the Universidad
Rey Juan Carlos.

\end{document}